# The influence of controlled vibration effects on fluid flow

Alexey Fedyushkin
*Laboratory of Complex Fluid Mechanics*
*Ishlinsky Institute for Problems in Mechanics of Russian*
*Academy of Sciences*
Moscow, Russia
fai@ipmnet.ru

*Abstract*. **This article presents the results of studies demonstrating the influence of nonlinear effects of laminar flow under vibrational harmonic effects on fluid flow and heat transfer. The paper summarizes the results of research on the influence of vibrations in various fluid flow problems. The effect of periodic oscillations on the symmetrization of an asymmetric flow in a diffuser, on Rayleigh-Bernard convection and on the wide of boundary layers in various single crystal growth processes are shown.**

## I. Introduction

During vibrational action on continuous media, their anomalous nonlinear peculiarities and resonant properties may manifest themselves [1-4]. Nonlinear peculiarities of the moving under vibration action are manifested not only in liquids, but also in the movement of bulk granular media [5]. The study of the effects of vibrations on liquid media has been carried out since the works of M. Faraday (1831) [6] and L. Rayleigh (1883) [7]. The vibrations in mechanical systems can be both a negative and a positive factor affecting the course of many physical, chemical, technical and technological processes. This article does not consider the negative effects of vibration on hydrodynamics and heat and mass transfer. Vibration effects can be an effective control tool for various processes. The vibrations are used in the processes of separation, transportation and mixing of various homogeneous non-isothermal, heterogeneous liquid and solid bulk media. Vibration effects abound in diversity, for example, vibration effects can be: 1. Translational (linear, circular, noncircular etc.), 2. Rotary (circular and noncircular), 3. Swinging, 4. The vibrations of all volume of a fluid or vibrations of borders or body immersed in a fluid, 5. Harmonic and nonharmonic, 6. Vibration with Low or high frequency and -small or long amplitude, 7. ACRT-accelerated crucible rotation technique [8, 9], etc. Vibrational control of the heat exchange in the melt is more energy-efficient and simpler than controlling the melt flow by changing the gravitational or magnetic field. Therefore, the study of vibration effects on the hydrodynamics of the melt is an actual task. Reviews of works on vibrational convective flow can be found in [9, 10]. The vibration effect on a liquid volume can be created in two ways: the first is the vibration of the entire liquid volume as a whole, it is the so–called g-jitter, and the second method is the case when vibrations are applied to a part of the boundary of the liquid volume or to a vibrator immersed in liquid. The first method of vibration exposure can be considered as a special case of the second case. Many theoretical papers [11-13] and experimental papers [14-18] have been devoted to the study of vibrations of the first type. The authors of books [11-13] for the first time pointed out the occurrence of an averaged vibrational-convective flow under periodic action on a liquid.

This paper presents and summarizes the results of mathematical modeling of the following problems: on flow symmetrization in a flat diffuser, on Rayleigh-Benard convection, and on the hydrodynamics of melt and heat and mass transfer in the processes of growing single crystals under vibration in relation to three methods of growing single crystals: Bridgman, Chokhralsky and floating zone [19-31]. The results are analyzed for quasi-stationary modes, and both instantaneous and time-averaged flow characteristics are presented. The results of numerical modeling have shown that vibrations can reduce the thickness of dynamic and temperature boundary layers and increase the temperature gradient at the crystallization front, which can intensify heat and mass transfer and, in particular, the rate of crystal growth [19-28]. The fact of increasing the crystal growth rate up to four times under vibrational action on the crystal was discovered experimentally in [14], which is an experimental confirmation of an increase in the temperature gradient at the crystallization front. The paper [29] shows the change in the beginning time and in structure of Rayleigh-Benard convection under vertical vibrations in a long-confined layer heated from below (the Rayleigh-Benard problem with vibration of the heated wall). This is important for the tasks of intensifying the cooling of devices and devices, as well as for the processes of metal melting and epitaxial crystal growth.

The study of the problem symmetrization of asymmetric fluid flows by means of vibration action on the flow is also important in a lot off applications, for example, in mechanical engineering for fuel injection in engines, as well as in biomedicine when creating new technologies and methods for the precise targeted delivery of drugs to the necessary areas of organs in human treatment. The effect of periodic disturbances on the input flow in a short mini diffuser was considered in [32], which numerically shows the effect of the solution angle and diffuser elongation on the asymmetry of the flow in a flat diffuser and told that by applying periodic vibrations to the input flow, the flow can be symmetrized, but this needs more detailed study. This paper presents the results on the symmetrization of the flow in a flat diffuser using two methods of vibration action [31].

The effect of vibrations on boundary layers is important not only in the crystallization of single crystals, but is also of great importance in other processes, in particular, in the processes of cooling and boiling [30]. The problem of the effect of vibrations on mass transfer is not considered in this paper, but the results on the mass transfer of impurities during crystal growth are presented in [23-27]. Also, a review of the

This work was supported by the Russian Science Foundation grant 24-29-00101.

works and results on the peculiarities of heat and mass transfer in boiling processes and under the influence of vibration were given in [30].

## II. MATHEMATICAL MODEL

The mathematical model is based on the numerical solution of a system of non-stationary planar 2D Navier-Stokes equations for natural convection of an incompressible liquid in the Boussinesq approximation (1-3):

$$\nabla \cdot \mathbf{u} = 0 \quad (1)$$
$$\rho_0 d\mathbf{u}/dt + \nabla p = \nabla \cdot (\rho_0 \nu \nabla \mathbf{u}) - \rho_0 g \beta (T - T_0) \mathbf{e}_z \quad (2)$$
$$\rho_0 c_V dT/dt = \nabla \cdot (k_T \nabla T) \quad (3)$$

where traditional notation is used. The problems were considered for flat cases or for conditions of axial symmetry with or without rotation. Therefore, for a cylindrical coordinate system $r, \theta, z$, then $u, v, w$ are radial, circumferential and axial velocity projections, $\nu, k_T$ are kinematic viscosity, heat conduction coefficients, $\beta$ is the buoyancy coefficient, $T_0$ is a reference temperature, $\rho_0$ is a reference density, g – acceleration of gravity opposite directed to the vertical coordinate axis (z). The boundary conditions were as follows: for velocity - no friction on a free surface, no slip condition on solid surfaces and setting the velocity of the vibrator or moving at the vibrating wall (on law y (or z) = $A sin(2\pi f t)$ with a frequency f and an amplitude A, $Re_{vibr} = A^2 2\pi f / \nu$ – is vibration Reynolds number); for temperature - were conditions of the first kind or thermal insulation conditions and at the crystallization interface, either the crystallization temperature or the Stefan condition with latent heat release was set.

The results presented in this paper were obtained using different numerical methods: the finite-difference scalar method [33], the fully implicit matrix finite-difference method [34, 35], and the conservative control volume method [36]. An algorithm which using for for solving the Stefan problem by the finite element method for modeling heat and mass transfer processes in a fluid with a phase transition was described in paper [37]. The good accuracy of numerical results was confirmed by comparison with experimental data and comparison of numerical results obtained by various numerical models.

## III. SYMMETRIZATION OF LAMINAR VISCOUS FLUID FLOW IN A FLAT DIFFUSER BY VIBRATIONAL EFFECT

The problem of the flow of a viscous incompressible liquid in a flat diffuser in the approximation of flow symmetry was solved by the authors of [38, 39]. It is known that when the Reynolds number increases above the critical Re* number, the flow loses symmetry, staying steady state and laminar. [40-46].

This article shows two methods of symmetrization of the asymmetric flow of a viscous incompressible liquid in a flat diffuser using periodic vibration action: 1 - from the side of the input stream, 2 - from the side of the walls of the diffuser. The research was carried out on the basis of solving the complete two-dimensional Navier-Stokes equations for an incompressible fluid (1, 2) for case g=0. The harmonic effects of vibration effects (in the form of $A \sin(2\pi f t)$, where A and f are the amplitude and the frequency of the changing velocity) on velocity are considered.

### A. The problem statement

The laminar flow of a viscous incompressible fluid driven through a channel bounded by two flat walls inclined towards each other at a small angle β is considered. In this paper we consider flat diffuser bounded by two arcs ("input" and "output" boundary) with the one center (Fig. 1a).

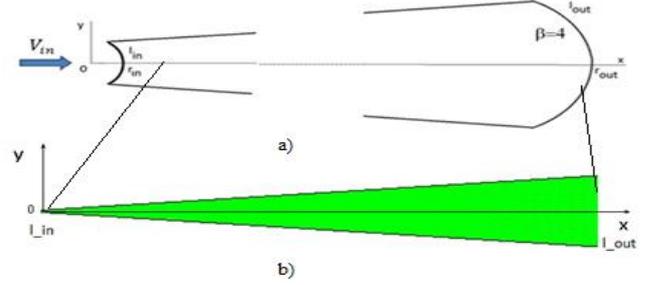

Fig. 1. Scheme of the computational domain for a flat diffuser: a) it is details of domain near the inlet and outlet of the diffuser; b) the numerical region with mesh ($\beta = 4°, L = 0.495$ m).

The geometry of the mathematical model was chosen in order to be able to compare our results with the results of well-known works [38, 39, 41 – 43]. Geometric model of the diffuser is as follows: opening angle is $\beta = 4°$, the input boundary has the form of an arc $l_{in}$ ($r_{in}$=0.005 m) where $r$ is calculated by formula $r^2 = x^2 + y^2$ (Fig. 1). The initial conditions are $t = t_0 = 0$, $V(t_0) = 0$, $P = 0$. The velocity scale is chosen by the velocity $V_{in}$ and the Reynolds numbers are defined as $Re = Re_{in} = V_{in} l_{in} / \nu$, $Re_{vibr} = A l_{in} / \nu$, $y_{dimless} = y/r \sin(\beta/2)$, $V_{x\_dimless} = V_x / V_{x\_in}$, $V_{y\_dimless} = V_y / V_{x\_in}$.

### B. The fluid flows in the diffuser without vibrations

The results for the case asymmetric fluid flows (Re=279) in the diffuser without vibration effects are presented in Fig. 2 [40]. The results coincide with results of papers [42, 43].

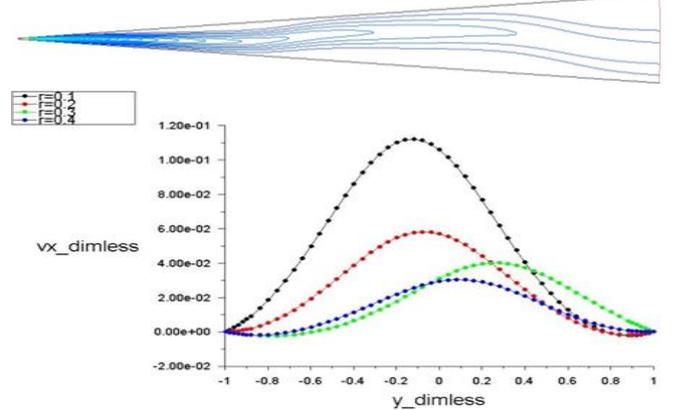

Fig. 2. The isolines and the profiles in vertical cross- sections of horizontal component $V_x$ of velocity vector for the case of asymmetrical steady state fluid flows (Re = 279).

*C. Only vibrational fluid flow in the diffuser*

In Fig. 3 the averaged profiles and isolines of the longitudinal velocity for the case of periodic velocity changes at the entrance to the diffuser $V = V_{in} + A \sin(2\pi f)$ ($V_{in} = 0$, $A = 1\,m/s$, $f = 10\,Hz$, $Re_{vibr}$=349) are shown. The averaged velocity profiles have velocity maxima near the walls – this is the "Richardson ring effect" [47].

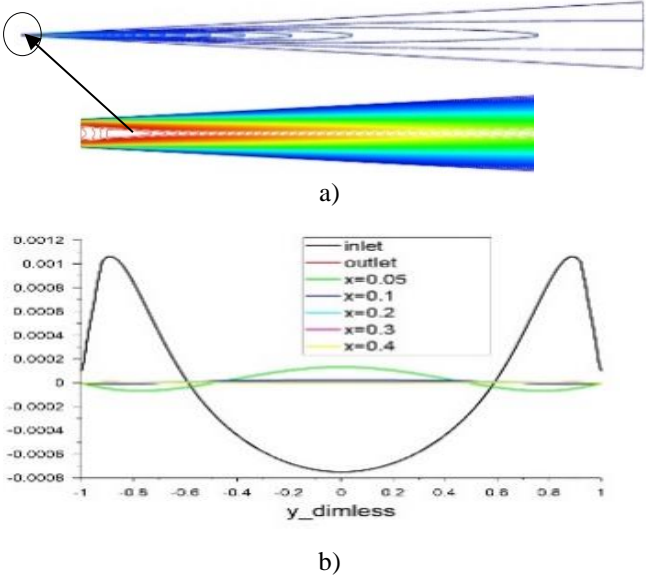

Fig. 3. The isolines of the averaged longitudinal component of the velocity mean_$V_x$, (below are the isolines of the mean_$V_x$ velocity near the entrance to the diffuser) (a), the profiles of the longitudinal component mean_$V_x$ of the velocity (b) for case $V_{in} = 0$, $A = 1\,m/s$, $f = 10\,Hz$

*D. Symmetrization of the fluid flow in the diffuser due to the effect of vibration on the inlet velocity*

The effect of a periodic vibrational disturbance $V = V_{in} + A \sin(2\pi f)$ ($f = 10\,Hz$, $A=0.1\,m/s$ $Re_{vibr}$=2.4) on the basic flow with Re=279 are presented in Fig. 4.

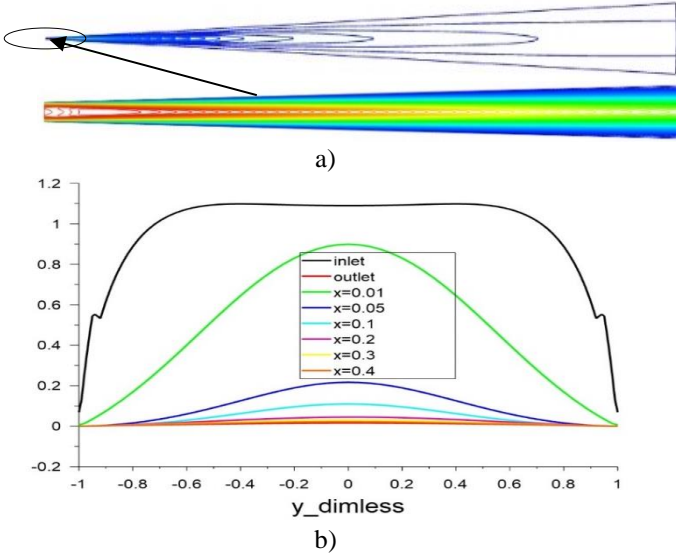

Fig. 4 The isolines of the averaged longitudinal component of the velocity mean_$V_x$, (below are the isolines of the mean_$V_x$ velocity near the entrance to the diffuser) (a), the profiles of the longitudinal component of the mean_$V_x$ velocity (b) for case $V_{in} = 11.7\,m/s$, $A = 0.1\,m/s$, $f = 10\,Hz$ (Re=279, $Re_{vibr}$=2.4)

Comparison of the results in Fig.2 and Fig. 4 shows that the effect of vibrations ($Re_{vibr}$=2.4), even at amplitudes less than 1% of the velocity $V_{in}$ (Re=279) can lead to symmetrization of the fluid flow in the diffuser.

*E. Symmetrization of the fluid flow in the diffuser due to the effect of vibration from the walls*

An example second approach of symmetrization of the fluid flow velocity in a flat diffuser by vibration action along normal to the walls of the diffuser according to the harmonic law $V_n = A \sin(2\pi f)$ with a small amplitude A and a frequency f is shown in Fig. 5. In Fig. 5 mean_$V_x$ – is the time-average velocity profiles for Re=279, A=0.001m/s, f=10 Hz ($Re_{vibr}$=0.02) are shown.

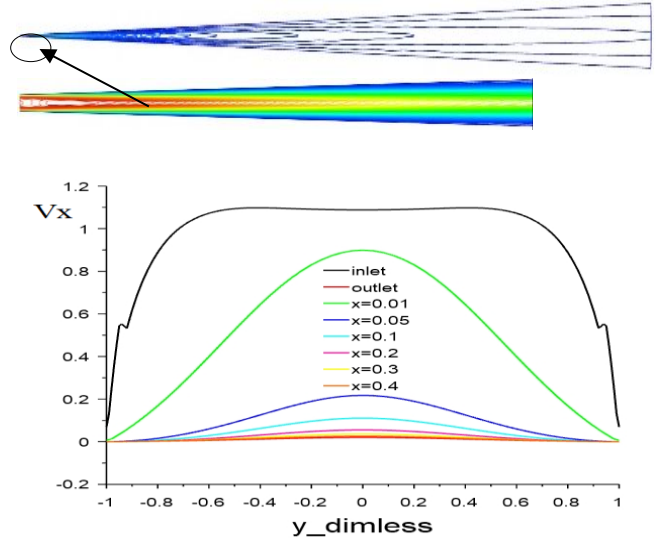

Fig. 5. The profiles of time average velocity (mean_$V_x$) for fluid flow in a flat diffuser with vibration action from the walls of the diffuser for Re=279, A=0.001m/s, f=10Hz ($Re_{vibr}$=0.02).

The results of numerical simulation have shown two ways of symmetrization of asymmetric laminar flows of viscous incompressible fluid in a flat diffuser: the first - due to a weak periodic effect on the flow velocity at the entrance to the diffuser and the second – due to vibration action from the walls of the diffuser. It is shown that the impact of vibration, even at amplitudes less than 1% of the velocity $V_{in}$, can lead to the symmetrization of the fluid flow in the diffuser. Richardson's "ring effect" (that is, the effect of the influence of harmonic oscillations of the input flow on the shape of the fluid flow velocity profile in a cylindrical pipe) was demonstrated for a diffuser.

IV. THE EFFECT OF CONTROLLED VIBRATIONS ON RAYLEIGH-BENARD CONVECTION

The problem of convective flow in a horizontal layer heated from below is called the Rayleigh-Benard (R-B) problem. This problem has a threshold character of the occurrence of natural convection, which is determined by the critical Rayleigh number.

R-B problem was considered for a horizontal layer with free top boundary with an aspect ratio of 1:10 and the Prandtl number Pr=1 in a gravity field with specified temperatures on horizontal walls and with thermally insulated vertical walls.

The analysis of solutions was carried out for a steady-state mode (or for the presence of vibrations, on a quasi-stationary mode).

The results of numerical simulation presented in Fig. 6 show the influence of the lower horizontal wall oscillations on the structure of the convective flow in the Rayleigh-Benard problem. The number of Rayleigh-Benard rollers decreases from 10 to 9 during vertical harmonic vibrations of the lower wall (on law $y = A sin(2\pi ft)$ with a frequency f=10 Hz and an amplitude A=$10^{-4}$ m, $Re_{vibr} = A^2 2\pi f / \nu = 0.007$), which indicates a decrease in the wave number of the periodic convective structure.

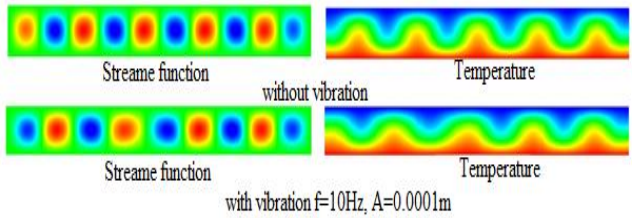

Fig. 6. Pictures of isolines of the stream function and isotherms with and without vibrations of the lower wall with $Re_{vibr} = 0.007$, $Ra = 4 \cdot 10^3, Pr = 1$.

The simulation results also showed the possibility of a significant decrease in the critical Rayleigh number for the occurrence of R-B convection under vibration action. The time of occurrence and establishment of the quasi-stationary regime of convective flow is also significantly reduced, as shown in Fig. 7.

A decrease in the critical Rayleigh number and the time of occurrence of Rayleigh-Benard convection due to vertical vibrations of the lower wall was also shown in paper [29].

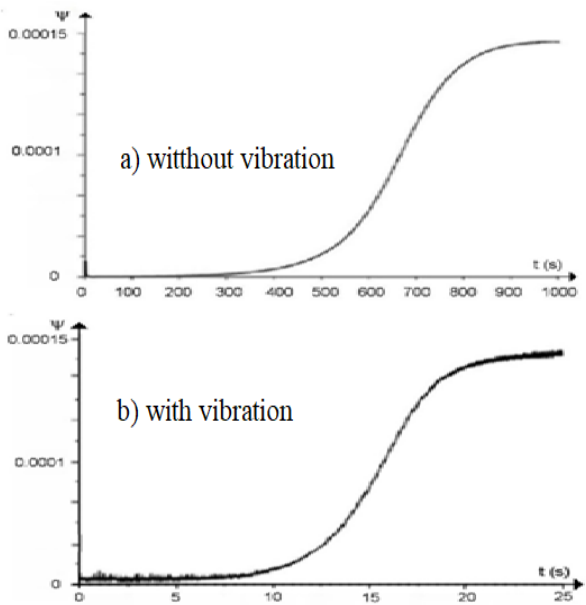

Fig. 7. The dependences of the maximum values of the stream function on time ($Ra = 4 \cdot 10^3, Pr = 1$): a) – without vibrations; b) – with vertical vibrations of the bottom wall with $Re_{vibr} = 0.007$.

## V. THE EFFECT OF VIBRATIONS IN CRYSTAL GROWTH PROCESSES

### A. Bridgman model

The calculation results were carried out for the following geometric configurations of crucibles for Bridgman method with submersible vibrators for a fixed flat and variable calculated shape of the crystallization front shown in Fig. 8. The area under consideration has the following dimensions: R=1.6 $10^{-2}$; H= 3.2 $10^{-2}$; $r_{vibr}$=4 $10^{-3}$; $h_1$=8 $10^{-3}$; $h_2$=8 $10^{-3}$; δ=$10^{-3}$ (m) where R is the radius of the ampoule, H is the height of the ampoule, h1 is the distance between the vibrator and the solid-liquid interface, h2 is the thickness of the vibrator (the distance between its lower and upper surfaces), the gap (the distance between the vibrator and the side wall of the crucible. The following variants with size values A=5 $10^{-4}$ and $10^{-4}$ m, f=0-100 (Hz) are calculated.

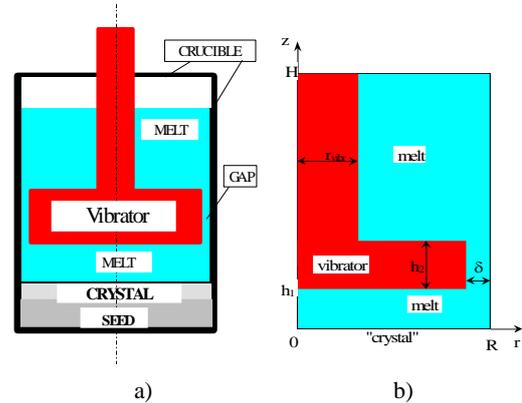

Fig. 8. The principal geometrical schemes for Bridgman crystal growth model with submerged vibrator, a) for Stefan problem with variable melt-crystal interface, b) model with fixed flat shape of the melt-crystal interface

*1) The effect of vibrations on temperature boundary layers*

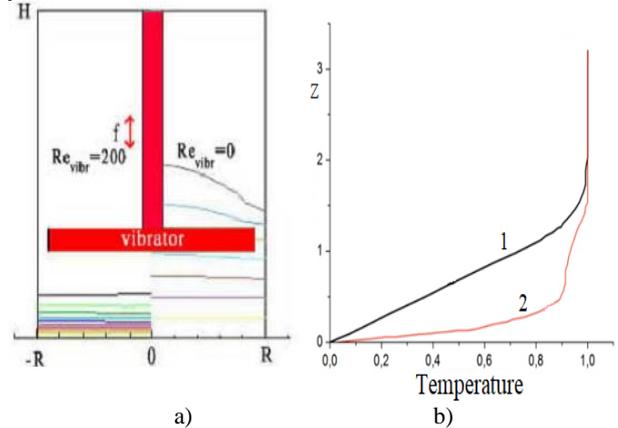

Fig. 9. a) - Isotherms in the NaNO$_3$ melt (Pr=5.43) (on the right – without vibrations, on the left – with vibrations Re$_{vibr}$=200), b) - vertical temperature profiles (r=0.75) in the NaNO$_3$ melt (Pr=5.43) (line 1 – without vibrations, 2 – with vibrations Re$_{vibr}$=200)

*2) The effect of vibrations on the shape of the crystallization front*

Using the method of solving the Stefan problem described in [37], for the Bridgman method with a submerged vibrator (Fig. 10), a simulation of convective heat transfer was performed in order to determine the effect of vibrations on the shape of the crystallization front.

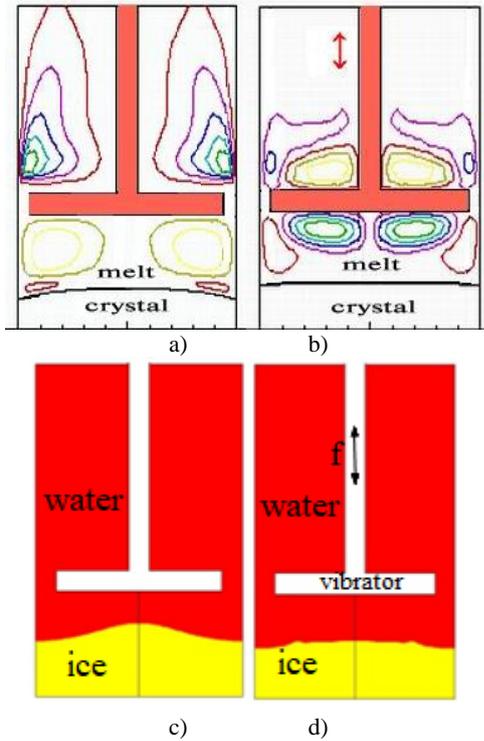

Fig. 10: The effect of vibrations on the shape of the front crystallization: a), b) – streame function in the melt of NaNO3, (a) without vibrations - f=0; b) - with vibrations A= $10^{-4}$ m, f=50 Hz), c), d) – water-ice interface, c) – without vibrations, f=0, d) – with vibrations, A= $10^{-4}$ m, f=30 Hz)

*B. Czochralski model with submerged vobrator*

The scheme of the computational domain is shown in Fig. 11. The computational domain is a square with sides L=H=3 cm Crystal with a diameter of d=1cm and immersed into the melt to a depth of 1mm, the vibrator has a diameter of 0.8 cm and thickness 1mm. It is assumed that the immersed vibrator is located under the crystal at a distance h horizontally and in the middle (parallel to the surface of the crystal). Irregular grids with refinement near the solid walls and the corners of the vibrator and the crystal were used in the calculations. The vibrator makes translational oscillatory movements along the vertical axis of the crystal according to the law: $y = y_0 + A\sin(2\pi ft)$, with frequency f and small amplitude A=$10^{-4}$ m, $y_0$ is initial location of vibrator.

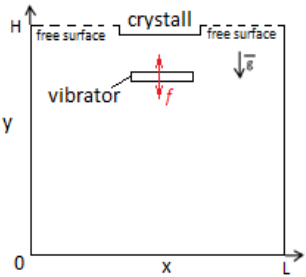

Fig. 11. Scheme of the computational domain.

The isotherm and structure of the averaged flow is presented in Fig. 12(a-c). It is show how the vibrating immersed activator leads to the mixing of the entire volume of the melt. In Fig. 12d presents temperature profiles on vertical cross section (on axis) that show the effect of vibration on the temperature boundary layer and the temperature gradient near the crystallization front (Pr=7; $Re_{vibr} = 1500$; h/d=0.5, A=4 $10^{-4}$ m, f=20Hz).

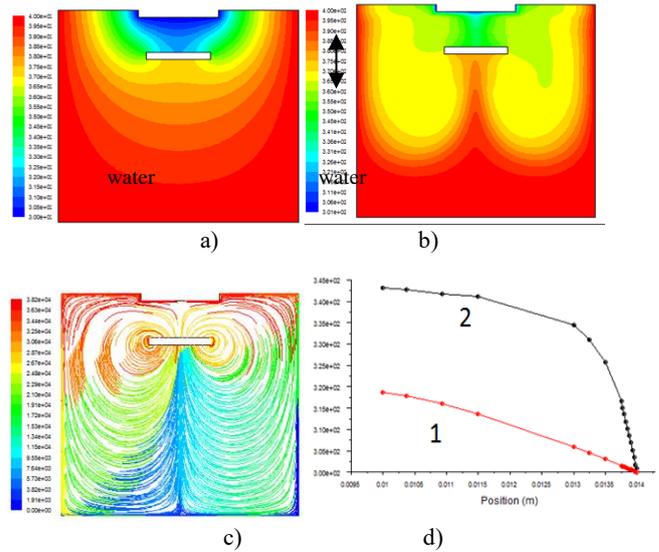

Fig. 12. Isotherms temperature: a) – without vibration, b) with vibrations c) flow tracks, d) temperature profiles on the axis section: curve 1- is without vibration, curve 2 - is with vibration. (Pr=7, $Re_{vibr} = 1500$, Ra=0).

*C. Floating zone model*

The scheme of both the design area and the boundary conditions for the zone melting model are shown in Fig. 13.

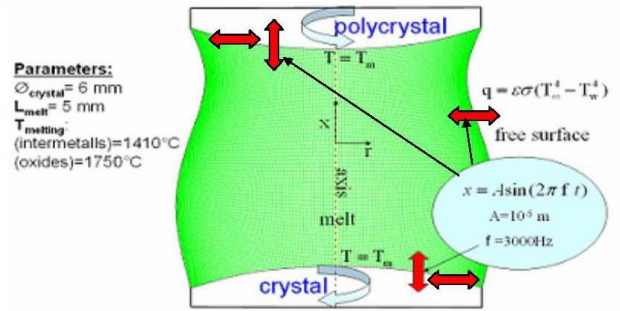

Fig. 13. The scheme calculation region for floating zone model

Figure 14 shows the results of calculating the hydrodynamics and temperature distribution during silicon crystallization by the zone melting method with fluctuations of the lower boundary (crystal). In this variant, the following factors were considered: natural and capillary convection, crystal rotation and counter-rotation of the polycrystal, radiation and vibrations from the crystal according to the harmonic law. Fig. 14 (c) shows a comparison of vertical temperature profiles for two cases with and without vibrations. From these results, it can be seen that the vibration effect reduces the temperature layer at the melt-crystal interface.

For the Bridgman, Czochralski and zone melting crystal growth methods, it is numerically shown that vibrations can reduce the width of the temperature boundary layer and, as a result, increase the temperature gradient at the melt-crystal interface (Fig. 9, 12, 14). An increase in the temperature gradient can intensify heat and mass transfer near the crystallization front, as well as the rate of crystal growth.

Calculations have shown that with a significant influence of vibrations on the boundary layer, they practically not affect temperature fluctuations at the melt-crystall interface . For example, for melts with a Prandtl number greater than one, temperature changes caused by vibration during one period of oscillation are no more than one percent.

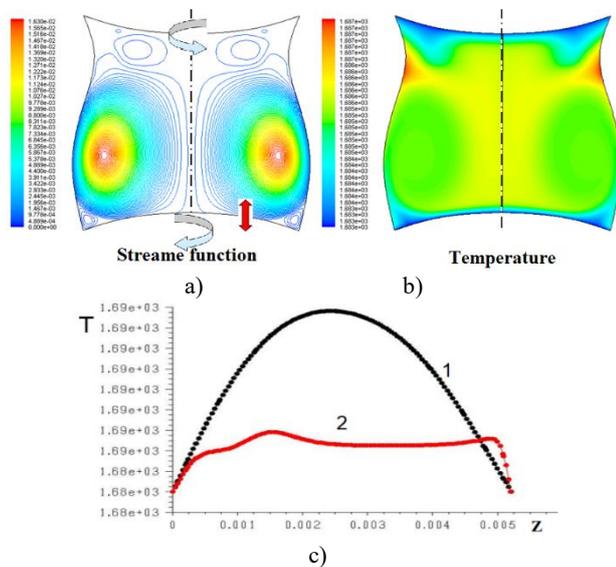

Fig. 14. a) – Streame function, b)- Isotherms in the melt Si (Pr=0.01) with natural and Marangoni convection, rotation, and vibrations. c) - vertical temperature profiles (r=0.75) (line 1 – without vibrations, 2 – with vibrations bottom wall ( A=$10^{-4}$ m , f=3kHz).

## VI. CONCLUSIONS

It is possible to symmetrize the flow of viscous liquid in the diffuser using a weak harmonic vibration effect from the inlet side or from the walls of the diffuser.

By controlled vibration action on the convective flow of the liquid, it is possible to reduce the thickness of the boundary layers, as well as to change the structure and time of occurrence of Rayleigh – Benard convection.

It has been found that vibrations can reduce the thickness of the boundary layer at the solid-liquid interface. It is shown that this effect occurs for three methods of growing single crystals: Bridgman, Chokhralski and zone melting. For the Bridgman model, it shows that it is possible to make the surface of the crystallization front flatter by means of vibration action. This is of fundamental importance for improving crystal growing technologies due to the possibility of using vibrations to control temperature gradients at the solid-liquid interface, i.e. to control the kinetics and rate of crystal growth.